\documentclass[preprint,showpacs,preprintnumbers,amsmath,amssymb]{revtex4-1}

\usepackage{graphicx}
\usepackage{dcolumn}
\usepackage{bm}
\usepackage[dvips]{color}

\begin{document}

\preprint{APS/123-QED}

\title{Photoionization spectroscopy of excited states of cold cesium dimers}

\author{Nadia Bouloufa$^1$}%

\author{Elena Favilla$^2$}

\author{Matthieu Viteau$^{1, 3}$}
\author{Amodsen Chotia$^{1, 3}$}%

\author{Andrea Fioretti$^{1, 2}$}
 \email{andrea.fioretti@lac.u-psud.fr}
\author{Carlo Gabbanini$^2$}

\author{Maria Allegrini$^{3, 1}$}%

\author{Mireille Aymar$^1$}%
\author{Daniel Comparat$^1$}%
\author{Olivier Dulieu$^1$}%
\author{Pierre Pillet$^1$}%
 \affiliation{(1) Laboratoire Aim\'e Cotton, B\^at. 505, CNRS, Universit\'e Paris-Sud 11, 91405 Orsay Cedex, France}%
 \affiliation{(2) Istituto Nazionale di Ottica, INO-CNR, U.O.S. Pisa "Adriano Gozzini", via Moruzzi 1, 56124 Pisa, Italy}%
 \affiliation{(3) CNISM, Dipartimento di Fisica, Universit\`a di Pisa, Largo B. Pontecorvo, 56127 Pisa, Italy.}

\date{\today}

\begin{abstract}
Photoionization spectroscopy of cold cesium dimers obtained by photoassociation of cold atoms in a magneto-optical trap is reported here. In particular, we report on the observation and on the spectroscopic analysis of all the excited states that have actually been used for efficient detection of cold molecules stabilized in the triplet $a^3\Sigma_u^+$ ground state. They are: the $(1)^3\Sigma_g^+$ state  connected to the $6s+6p$ asymptote, the $(2)^3\Sigma_g^+$ and $(2)^3\Pi_g$ states connected to the $6s+5d$ asymptote and finally the $(3)^3\Sigma_g^+$ state connected to the $6s+7s$ asymptote. The detection through these states spans a wide range of laser energies, from 8000 to 16500~cm$^{-1}$, obtained with different laser dyes and techniques. Information on the initial distribution of cold molecules among the different vibrational levels of the $a^3\Sigma_u^+$ ground state is also provided. This spectroscopic knowledge is important when conceiving  schemes for quantum manipulation, population transfer and optical detection of cold cesium molecules.
\end{abstract}

\maketitle

\section{Introduction}
\label{intro}

Cold and ultracold molecules have received increasing attention in recent years due to the significant advances and the potential new applications that they offer in several directions, like ultra-high resolution molecular spectroscopy, tests of fundamental theories in physics, clocks based on molecular transitions, full control of the dynamics of cold chemical reactions, design of quantum information devices~\cite{Smith2008, fried2009, Gabbanini-Dulieu2009,Carr2009}. Among the various approaches to form stable cold molecules, the photoassociation (PA) of laser-cooled atoms into an excited molecule followed by stabilization via spontaneous emission has proven very successful to obtain large and dense samples of ultracold homonuclear and heteronuclear alkali dimers~\cite{Gabbanini-Dulieu2009}. Ultracold molecules in the lowest rovibrational level of their electronic ground state have been observed in this way~\cite{1999PhRvL..82..703N, deiglmayr2008a}, or through subsequent vibrational cooling using a pulsed laser~\cite{2008Sci...321..232V, viteau:021402,2009NJP...submitted}. Spectacular developments in this direction have also been achieved through the control of Feshbach resonances of ultracold colliding alkali atoms with external magnetic fields combined with the Stimulated Raman Adiabatic Passage (STIRAP) method to create ultracold molecules in their absolute ground state level~\cite{ni2008,danzl2009}.

\begin{figure}
\centering
\includegraphics[width=0.6\textwidth]{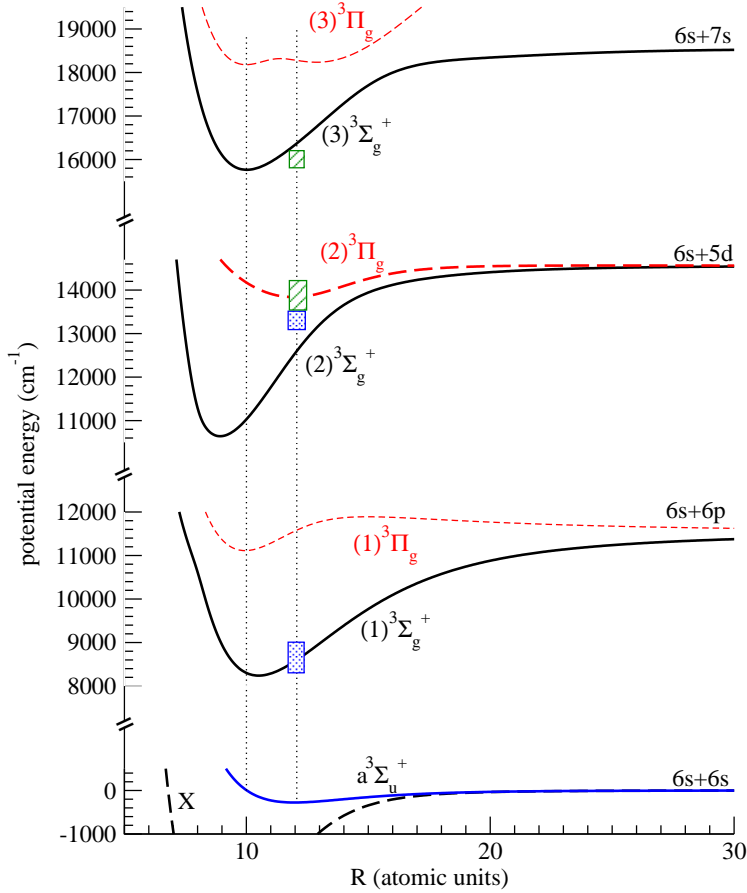}
\caption{(Color online) Potential curves relevant for the present study. The boxes indicate the energy regions made accessible by the REMPI lasers available in Pisa (blue shaded boxes) and in Orsay (green hatched boxes). The vertical dashed lines visualize the distance of the innermost turning point (left line) and of equilibrium distance (right line) of the $a^3\Sigma_u^+ (6s+6s)$. The potential curves are taken from ref.~\cite{aymar2010}. The $(1)^3\Sigma_g^+ (6s+6p)$ and $(1)^3\Pi_g (6s+6p)$ states combine together to form the $0_g^-(6s+6p_{3/2})$ state which is used for the PA step (see Fig.\ref{fig:PA_PES}).}
\label{fig:REMPI_PES}
\end{figure}

Most of these studies benefited from a very efficient method of detecting those molecules, through Resonant MultiPhoton Ionization (REMPI) into molecular ions~\cite{1998PhRvL..80.4402F} detected by time-of-flight. This method has also turned out to be a powerful way to explore the spectroscopy of molecular states of alkali dimers which were poorly known previously~\cite{dion2002,wang2005,wang2006,lozeille2006}. Here we report on the spectroscopy of all the excited states that have actually been used for efficient detection of ultracold Cesium molecules stabilized in the lowest metastable triplet state $a^3\Sigma_u^+ (6s+6s)$, namely the $(1)^3\Sigma_g^+ (6s+6p)$ state, the $(2)^3\Sigma_g^+ (6s+5d)$ and $(2)^3\Pi_g (6s+5d)$ states and the $(3)^3\Sigma_g^+ (6s+7s)$ state (see Figure~\ref{fig:REMPI_PES}). The formation process relies on PA of levels embedded in the double-well excited molecular state known as the $0_g^-(6s+6p_{3/2})$ state, which has been carefully analyzed recently~\cite{Bouloufa2007}. Detection via these states spans a wide range of REMPI laser energies, from 8000~cm$^{-1}$ to 16500~cm$^{-1}$, obtained with different laser dyes and techniques. The analysis combines REMPI data with all other experimental information available in the literature, as well as the most recent theoretical calculations performed by the theory group in Orsay. The knowledge of the population over vibrational levels of the $a^3\Sigma_u^+$ state of the created ultracold molecules is also extracted, clarifying in turn their formation mechanism during the PA step via a possible tunneling effect in the $0_g^-(6s+6p_{3/2})$ state~\cite{vatasescu2000,vatasescu2006}.

The article is organized as follows: in Sec.~\ref{exp_setups} we provide a brief description of the two experimental setups running in Orsay and Pisa, as well as the photoassociation step chosen in each case. In Sec.~\ref{exp_results} we present the experimental spectra discussing the spectroscopic information that can be extracted from them. Finally we present in Sec.~\ref{conclusions} conclusions and perspectives of the present work.


\section{Experimental setups in Orsay and in Pisa}
\label{exp_setups}

\begin{figure}
\centering
\includegraphics[width=0.6\textwidth]{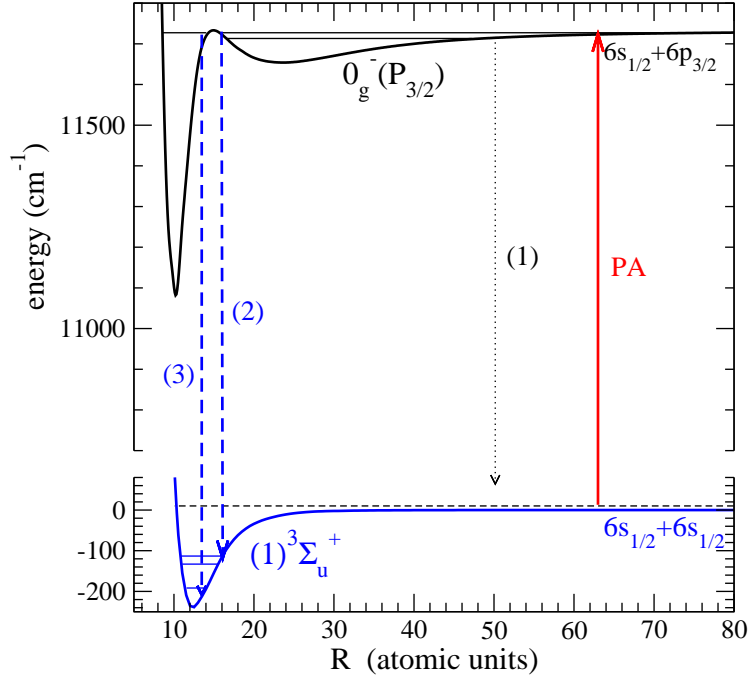}
\caption{(Color online) Scheme for the formation of deeply-bound ultracold Cs$_2$ molecules in the $a^3\Sigma_u^+ (6s+6s)$ state (arrows (2) and (3)), after PA of levels embedded in the outer well of the $0_{g}^{-}(P_{3/2})$ excited state (arrow labeled with PA). The possible destruction of the excited molecule by spontaneous emission into two free atoms is also indicated for completeness (arrow (1)).}
\label{fig:PA_PES}
\end{figure}

Observation of translationally cold cesium molecules has been obtained as a consequence of stabilization by spontaneous emission after the creation of an excited state molecule by PA below the $6s+6p_{3/2}$ asymptote~\cite{1998PhRvL..80.4402F} in a magneto-optical trap. When performing photoassociation, vibrational levels in a number of different electronic states can be excited, which subsequently lead to trap loss and possibly to cold molecule production (see figure \ref{fig:PA_PES}). In this way, PA spectroscopy of the long range part of these states can be performed~\cite{FiorettiEPJD1999,Comparat1999229, comparat2000, pichler:1796, pichler:6779, Ma2009106}. Depending on the shape of the excited state potential curve, the excitation can be followed preferentially by a spontaneous emission into two free atoms, or into rovibrational levels of the lowest stable electronic states of the dimer.

In Cesium, the $0_{g}^{-}(6s_{1/2}+6p_{3/2})$ state (hereafter referred to as $0_{g}^{-}(P_{3/2})$) exhibits a potential curve with a remarkable double-well shape, described in ref.~\cite{FiorettiEPJD1999}: besides the deep inner well located in the chemical range (with an equilibrium distance around 10~a.u.), a shallow outer well arises from the competition between the long-range electrostatic interaction between the $6s$ and the $6p$ atoms (varying as $1/R^3$, where $R$ is the internuclear distance) and the spin-orbit interaction (quite large in the $6p$ cesium state, \emph{i.e.} 554.1~cm$^{-1}$) which couples the $(1)^3\Sigma_g^+ (6s+6p)$ and the $(1)^3\Pi_g (6s+6p)$ states. To our knowledge there is no available spectroscopic information for the inner well which matches the $(1)^3\Pi_g$ state. In contrast, the outer well is well represented by an analytical asymptotic model~\cite{Bouloufa2007} which accurately accounts for all available spectroscopic data of ref.~\cite{FiorettiEPJD1999}. Due to the potential barrier between the two wells, the outer one presents a favorable Condon point at intermediate internuclear distances (around 15~a.u.) for spontaneous emission towards levels of the stable $a^3\Sigma_u^+ (6s+6s)$ state (see figure \ref{fig:PA_PES}). It is worthwhile to mention that among all alkali pairs, the Rb$_2$ molecule is the only one with the same pattern also leading to ultracold molecule formation~\cite{gabbanini2000,fioretti2001}.

The spectra reported in this paper have been obtained in two different experiments, one in Orsay and one in Pisa. They both rely on the same PA transitions towards the $0_{g}^{-}(P_{3/2})$ state. The two setups have similar characteristics: a magneto-optical trap (MOT) for cesium atoms, and a detection scheme for Cs$_2$ molecules by REMPI into Cs$_2^+$ ions observed using the time-of-flight technique, as a probe of the formation of ultracold Cesium dimers. Spectra of molecular ion yield are recorded by varying the frequency of the REMPI laser frequency. The two setups mainly differ by the laser equipment for both the PA and the detection steps, as detailed in the following subsections.

\subsection{The Orsay setup}
\label{Orsay_setup}

The core of the experiment is a Cs vapor-loaded MOT where nearly $10^8$ atoms are continuously laser-cooled and trapped at a temperature of $T\sim 100$~$\mu$K and at a peak density of $n\sim 10^{11}$~cm$^{-3}$. The MOT is created at the crossing of three pairs of counter-propagating laser beams, and at the zero-field point of a quadrupolar magnetic field. The cooling laser is provided by a master-slave diode laser system, where the  master laser is a 50~mW Distributed Bragg Reflector laser (SDL 5702 H1) and the slave one is a 150~mW laser (SDL 5422 H1). Its frequency is tuned 12.5~MHz to the red of the $6s(F=4)\rightarrow 6p_{3/2}(F'=5)$ cycling transition, where $F$ represents the hyperfine quantum number. An additional repumping laser, DBR SDL 5712 H, with a frequency tuned to the $6s(F=3)\rightarrow 6p_{3/2}(F'=4)$ prevents atoms from being optically pumped into the lower $F=3$ hyperfine state, thus exiting the cooling cycle. PA is achieved using a cw Titanium:Sapphire
laser (intensity 300~W~cm$^{-2}$) pumped by an Ar$^+$ laser. The frequency of the PA laser is red-detuned with respect to the $6s\rightarrow 6p_{3/2}$ atomic transition, in order to reach the known levels of the $0_{g}^{-}(P_{3/2})$ state. If ultracold stable molecules are formed in the lowest triplet state $a^{3}\Sigma _{u}^{+}$, they remain available for further studies during about 10~ms before they ballistically leave the MOT region. Their detection is achieved by REMPI using a pulsed dye laser (wavenumber $\sim$13500-16000~cm$^{-1}$, spectral bandwidth 0.1~cm$^{-1}$) pumped by the second harmonic of a pulsed Nd:YAG laser (repetition rate 10~Hz, duration 7~ns). Two metallic grids placed around the MOT create an electric field inside the MOT region so that the formed Cs$_{2}^{+}$ ions are quickly extracted, and detected by time-of-flight using a pair of microchannel plates. Further details on the experimental setup and methods can be found in ref.~\cite{Viteau2009} and references therein. The frequencies of both the PA and REMPI lasers are measured with a high finesse spectrometer, Ansgtrom WS8, with up to 15~MHz resolution in the case of a cw laser. The output of this spectrometer, which is recorded together with ion yield, provides the energy scale of the Orsay spectra.

In the present study two ranges of REMPI laser energies are explored, namely the 13500-14300~cm$^{-1}$ region, and the 15800-16200~cm$^{-1}$ region marked with hatched green boxes in Fig.~\ref{fig:REMPI_PES}. The level structure of the $(2)^3\Sigma_g^+(6s+5d)$, $(2)^3\Pi_g(6s+5d)$, and $(3)^3\Sigma_g^+(6s+7s)$ states is expected to be attainable in these ranges, and has never been reported elsewhere except for the $(2)^3\Pi_g(6s+5d)$~\cite{diemer1991, Kim1993}. This represent an important outcome of the molecular PA spectroscopy already pointed out in the past~\cite{stwalley1999}: stable alkali molecules are easily created in the triplet multiplicity in contrast with the situation in thermal gases where they mainly exist in their singlet ground state. Therefore the access is opened here to the triplet manifold of these species. It is worthwhile to mention here that the molecular spectroscopy of alkali dimers performed on helium droplets also offers such a possibility, as illustrated for instance in ref.~\cite{ernst2006} where the position of the absorption band of the $a^3\Sigma_u^+ (6s+6s)$-$(3)^3\Sigma_g^+(6s+7s)$ transition has been located around 16700~cm$^{-1}$.

\subsection{The Pisa setup}
\label{Pisa_setup}

Starting from a Cs vapor cell made in stainless steel with several viewports, a MOT is produced by three couples of orthogonal retroreflected laser beams and a quadrupole magnetic field. The cooling laser is a DFB diode laser (150~mW power), with a frequency tuned two line widths $\Gamma(=2\pi \times 5.22$~MHz) below the $6s_{1/2}(F=4)\rightarrow 6p_{3/2}(F=5)$ transition, i.e. at a similar detuning as that in the Orsay experiment. Another diode laser in an external cavity configuration, tuned on the $6s_{1/2}(F=3)\rightarrow 6p_{3/2}(F=4)$ transition acts as a repumper. Both lasers are frequency-locked in separated cells through saturated absorption spectroscopy. The MOT captures nearly 10$^{7}$ Cs atoms at a density of about 10$^{10}$~cm$^{-3}$. The PA step is performed with a dedicated 150~mW DFB diode laser also locked to a stable external cavity.

The dimers are first resonantly excited by an infrared photon pulse and ionized by a green photon pulse afterwards. The green pulse is obtained by extracting a fraction (about 1~mJ energy) of the second harmonic of a Nd:YAG laser (Continuum Surelite I-20, with 20 Hz repetition time), while the largest fraction of the beam pumps a dye laser (Quantel TDL50) operating with various combination of Rhodamine dyes (6G, 610, 640), in order to cover a sufficiently wide spectral region in the visible range. In order to obtain the IR radiation, the dye laser beam is focused into a 70~cm metal cell filled with hydrogen at high pressure (30~bar), where stimulated Raman scattering occurs. For vibrational scattering, molecular hydrogen combines a good conversion efficiency with a large energy shift (4155.1~cm$^{-1}$). Visible radiation can be converted into IR light using either the 1$^{\rm st}$ or the 2$^{\rm nd}$ Stokes emission. The dye laser beam is firstly focused by a 25~cm focal lens placed near the entrance window of the Raman cell, refocused by a second lens with 10~cm focal length, placed inside the cell, and finally a third lens outside the cell recollimates the beam. The 2$^{\rm nd}$ Stokes beam, which is separated from the pump and first Stokes beams using optical filters, has a maximum energy of about 100~$\mu$J. The IR and green beams are softly focused onto the cold Cs sample, impinging almost collinearly after independent optical paths. The IR pulse, having 5~ns width, arrives a few ns before the green pulse (7~ns width). The produced atomic and molecular ions are repelled by a grid, separated by time-of-flight and detected by a microchannel plate. The ion signals are recorded by gated integrators. With the choice of laser dyes above, the 8000-9500~cm$^{-1}$ and 12500-13500~cm$^{-1}$ spectral ranges can be covered for the first REMPI step, exciting the $(1)^3\Sigma_g^+(6s+6p)$ and $(2)^3\Sigma_g^+(6s+5d)$ molecular states (see shaded blue boxes in Figure~\ref{fig:REMPI_PES}). As mentioned above, no other experimental information is available on the $(2)^3\Sigma_g^+(6s+5d)$ states in the literature, while the structure of the $(1)^3\Sigma_g^+(6s+6p)$ state has been previously studied by Fourier Transform Spectroscopy by Amiot and Verg\`es~\cite{Amiot85}, and a few levels of the same state have been assigned in recent ultracold molecule experiment in Innsbruck~\cite{mark2009}.

\section{Analysis of the experimental REMPI spectra}
\label{exp_results}

\begin{figure}
\centering
\includegraphics[width=0.6\textwidth]{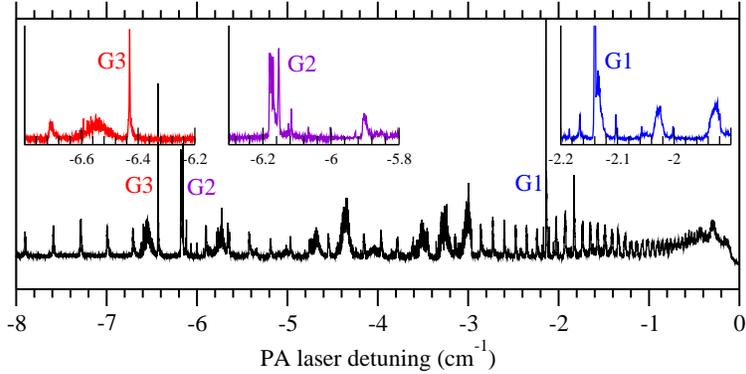}
\caption{(Color online) Cs$_2^+$ yield as a function of the detuning of the photoassociation laser frequency from the $6s_{1/2}(F=4)\rightarrow 6p_{3/2}(F=5)$ transition frequency. The region of the $0_{g}^{-}(P_{3/2})$ photoassociation spectrum exhibiting the giant  $G1$, $G2$, and $G3$ lines is displayed.}
\label{fig:0gmPAspectrum}
\end{figure}

As it has been demonstrated by Fioretti {\it et al.}~\cite{FiorettiEPJD1999}, three PA transitions producing three intense lines in the Cs$_2^+$ spectra - hereafter referred to as giant $G1$, $G2$, and $G3$ lines - have been observed, as a manifestation of an enhanced formation rate of ultracold Cs$_2$ molecules. These lines are located close to three PA lines assigned to vibrational levels $v=103$ , $v=80$, and $v=79$ of the $0_{g}^{-}(P_{3/2})$ outer well for the $G1$, $G2$, and $G3$ lines respectively. We reproduce the related portion of the PA spectrum in Figure~\ref{fig:0gmPAspectrum}. As shown in the insets, the $G1$ and $G2$ lines have a pronounced rotational structure which strongly differs from the broad unstructured neighboring lines. In contrast the $G3$ line is narrow and does not display any substructure. We focus on these lines in the present analysis as the corresponding excited levels are predicted to decay by spontaneous emission towards deeply-bound rovibrational levels of the $a^3\Sigma_u^+ (6s+6s)$ state. Indeed, Vatasescu \emph{et al.}~\cite{vatasescu2000,vatasescu2006} interpreted the large rotational structure of the $G1$ and $G2$ lines as being due to the resonant coupling of the $v=103$ and $v=80$ levels of the outer well with levels of the inner well due to tunneling through the potential barrier. Therefore the corresponding radial wave functions actually explores the inner well, which is then suitable for efficient spontaneous decay down to deeply-bound $a^3\Sigma_u^+ (6s+6s)$ levels (arrow (2) in Fig.~\ref{fig:PA_PES}). Using available theoretical potential curves, Vatasescu \emph{et al.}~\cite{vatasescu2006} predicted that the population of the $a^3\Sigma_u^+$ vibrational levels should be mainly spread over a narrow range around $v"=5,6$. The $G3$ structure remained unexplained at that time. Therefore one goal of the present analysis is to check if the recorded REMPI spectra can be interpreted in this manner. As will be seen in the following paragraphs, the present results actually suggest that the $G1$ and $G3$ lines are due to the mechanism proposed in refs.~\cite{vatasescu2000,vatasescu2006}, while the origin of the $G2$ line now remains unexplained. Indeed, all the recorded spectra show very similar structures when the PA laser is fixed on the $G1$ and $G3$ lines, while almost no structure is visible in the energy ranges used for ionization in the present paper when the PA laser is tuned on the $G2$ resonance. This is the reason why no REMPI spectra with PA on the $G2$ line are shown.

All the interpreted spectra concern excitation of triplet excited state manifolds starting from the $a^3\Sigma_u^+ (6s+6s)$ ground state. We used two determinations for the potential curve of this state that we will refer in the following as theoretical and experimental potential curves respectively. First the theoretical potential curve has been built as follows: the inner part ($R<20$ a.u.) is taken from ref.~\cite{foucrault1992}, and is matched to the accurate asymptotic part derived in  ref.~\cite{vanhaecke2004}. The short-range repulsive wall of the potential curve is slightly adjusted in order to reproduce the experimental triplet scattering length of ref.~\cite{chin2004}. The second potential curve, the experimental one, has been determined by Xie {\it et al.}~\cite{xie:051102} from accurate spectroscopic investigations. The equilibrium distance and harmonic constant of the theoretical $a^3\Sigma_u^+$ curve are equal respectively to 11.97a.u., and to 10.7~cm$^{-1}$, in reasonable agreement with the experimental values 12.32a.u. and 11.6~cm$^{-1}$~\cite{xie:051102}. The potential well is found 40~cm$^{-1}$ deeper in ref.~\cite{xie:051102} than in the calculation of ref.~\cite{aymar2010}.

Vibrational energies and wave functions for this state, as well as for the excited states, are computed with the Mapped Fourier Grid Representation (MFGR) method~\cite{kokoouline1999}. In a second step, theoretical line positions are evaluated from these vibrational energies, and possibly shifted by a constant energy to better match the experimental spectra. An estimate of the line intensities is obtained from the overlap of computed vibrational wave functions without involving transition dipole moments. The detailed analysis of the intensity patterns is not relevant here, as most of the spectra are recorded with strongly saturated transitions. In the following, the observed transitions will be identified only with vibrational indexes as their rotational structure cannot be resolved. Indeed, the accuracy of the energy positions is limited by the spectral width of the REMPI lasers and by power broadening leading to a resolution of $\approx \pm 0.5$~cm$^{-1}$ for the Orsay spectra. As detailed in the following sections, the resolution of the Pisa spectra is slightly worse (from $\approx \pm 1.0$~cm$^{-1}$ to $\approx \pm 2.0$~cm$^{-1}$) and depends on the laser used for the REMPI step.

\begin{figure}
\centering
\includegraphics[angle=270, width=0.6\textwidth]{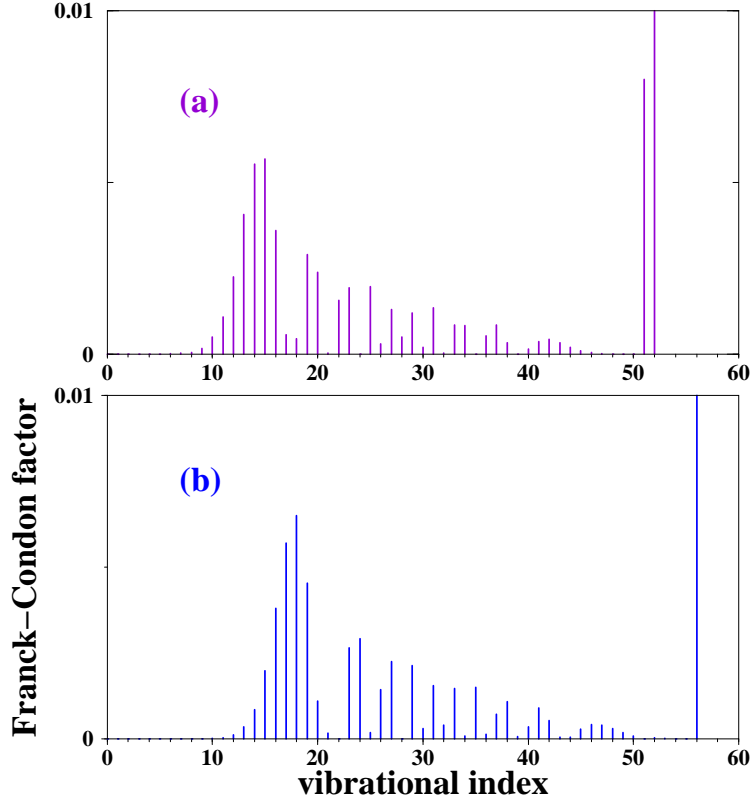}
\caption{(Color online) Overlap of the $v=79$ vibrational wave function of the $0_{g}^{-}(P_{3/2})$ potential curve, simulating the $G3$ resonance, with the vibrational levels of the $a^3\Sigma_u^+$ state (in abscissae). (a) using the theoretical $a^3\Sigma_u^+$ potential curve (see text); (b) using the experimental  $a^3\Sigma_u^+$ potential curve of ref.~\cite{xie:051102}.}
\label{fig:vibpop_astate}
\end{figure}

Prior to the interpretation of the experimental data, we first refined the prediction for the population distribution over $a^3\Sigma_u^+$ vibrational levels using the most recent determination of the $0_{g}^{-}(P_{3/2})$ external well~\cite{Bouloufa2007} which is smoothly matched for $R<15$a.u. to the $(1)^3\Pi_g (6s+6p)$ potential curve from ref.~\cite{aymar2010}. The potential barrier between these two wells is chosen as recommended in the model of ref.~\cite{vatasescu2006} with its top lying 2~cm$^{-1}$ above the energy of the $6s+6p_{3/2}$ dissociation limit. We computed the vibrational wavefunction of the level labeled as $v=79$ of the $0_{g}^{-}(P_{3/2})$ external well, which is resonant with a level of the internal well and coupled to it by tunneling, therefore simulating the $G3$ resonance. Then we calculate the overlap integral of this wavefunction with all the  $a^3\Sigma_u^+$ vibrational wavefunctions. Fig.~\ref{fig:vibpop_astate}a and Fig.~\ref{fig:vibpop_astate}b show that the population is expected to be spread over a range of $a^3\Sigma_u^+$ vibrational levels, broader than initially predicted in ref.~\cite{vatasescu2006}. The vibrational levels in the $v"=11-16$ range of the theoretical curve (i.e. $v"=15-20$ in the experimental numbering scheme) of the $a^3\Sigma_u^+$ state should be mostly populated. The large population of the $v"=51,52$ in Fig.~\ref{fig:vibpop_astate}a (resp. $v"=56$ in Fig.~\ref{fig:vibpop_astate}b) levels is due to the long-range part of the wave function located in the $0_{g}^{-}(P_{3/2})$ outer well.

\subsection{Ionization through the $(3)^3\Sigma_g^+(6s+7s)$ state}
\label{3_3_Sigma}

\begin{figure}
\centering
\includegraphics[width=10cm]{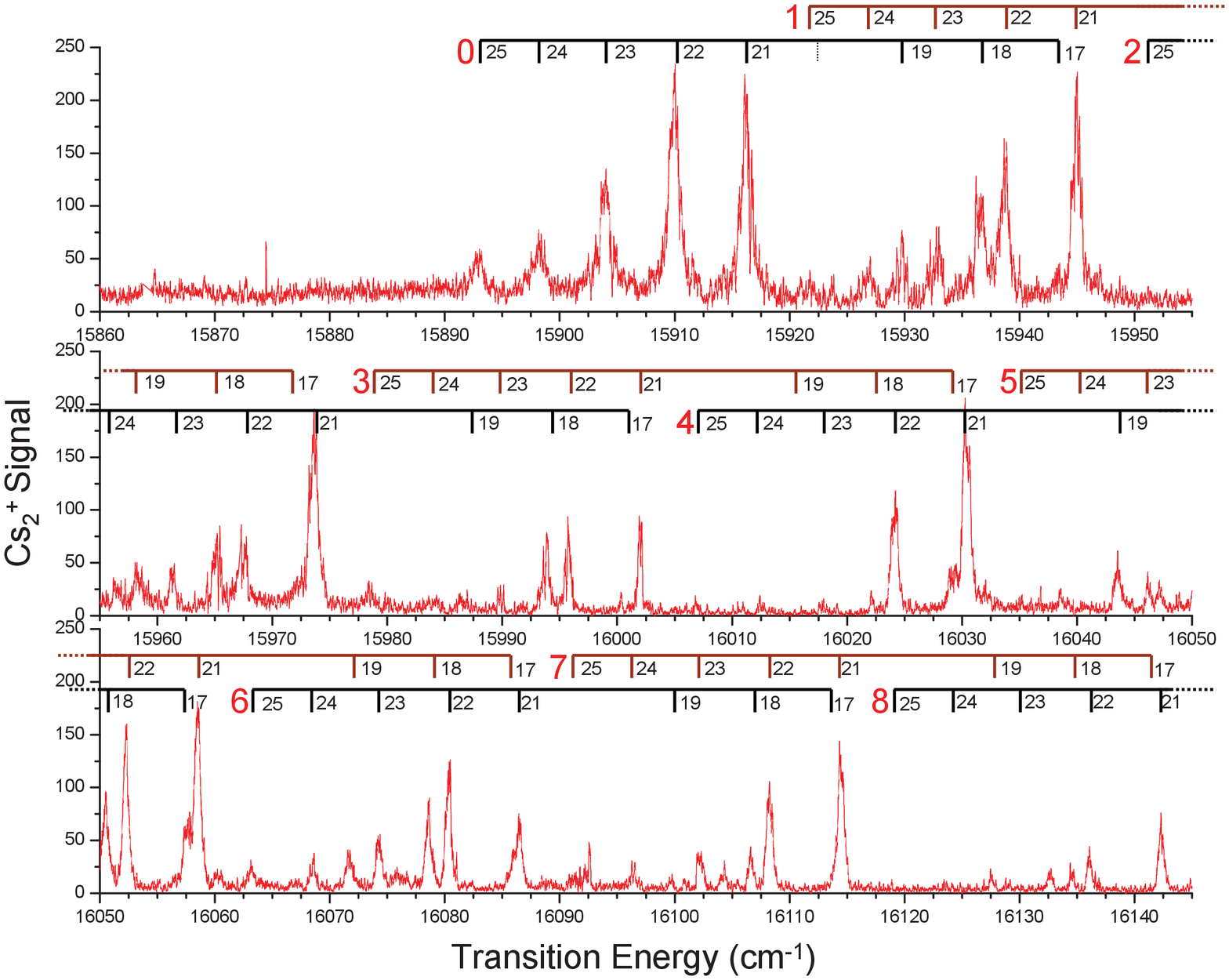}
\caption{(color online) Experimental REMPI spectrum recorded in Orsay between 15830 and 16140~cm$^{-1}$, when the PA laser is tuned on the $G1$ line. The transitions are assigned to $a^3\Sigma_u^+(v") \rightarrow (3)^3\Sigma_g^+ (v')$ transitions. Each series is labeled with $v'$ indexes (large and red) determined by the present experiment, and successive $v"$ indexes according to results of ref.~\cite{xie:051102}.}
\label{fig:3_3Sigma_G1}
\end{figure}

We start our analysis with the highest range of transition energies investigated here, namely between 15800 and 16200~cm$^{-1}$. As it can be seen in Fig.~\ref{fig:3_3Sigma_G1}, the recorded spectrum produced after PA of the $G1$ resonance is characterized by well isolated lines. This is indeed suggested by the somewhat isolated location of the $(3)^3\Sigma_g^+(6s+7s)$ potential curve (Fig.~\ref{fig:REMPI_PES}). The PA of the $G3$ resonance produces a spectrum (not displayed here) very similar to the one produced by the $G1$ line. As announced above, this is the first indication that the $G1$ and $G3$ are indeed induced by the same mechanism. We see that the same pattern is repeated along the whole spectrum, which facilitates its interpretation: as the theoretical harmonic vibrational constant of the $(3)^3\Sigma_g^+$, 27.2~cm$^{-1}$~\cite{aymar2010}, is predicted to be larger than for the $a^3\Sigma_u^+$ state (11.2~cm$^{-1}$~\cite{xie:051102}), successive series associated to the excitation of $(3)^3\Sigma_g^+$ levels are visible. The line recorded at the lowest excitation energy, 15893.0(5)~cm$^{-1}$, is assigned to the excitation of the $v'=0$ level of $(3)^3\Sigma_g^+$, which therefore fixes the absolute energy position of this previously unobserved potential curve. The splitting between the nine successive $(3)^3\Sigma_g^+$ vibrational levels observed is almost constant, it allowed us to deduce the harmonic constant of this state ($\approx 28.9 (5)$~cm$^{-1}$), in good agreement with the theoretical value of the harmonic constant above. We deduce the energy distance $T_e$ between the bottom of the $(3)^3\Sigma_g^+$ well and the experimental $a^3\Sigma_u^+$ well, 16092.8(5)~cm$^{-1}$, which would correspond to a shift of $14.6$~cm$^{-1}$ downwards of the bottom of the computed $(3)^3\Sigma_g^+$ curve relative to the $6s+6s$ asymptote.

\begin{figure}
\centering
\includegraphics[width=0.6\textwidth]{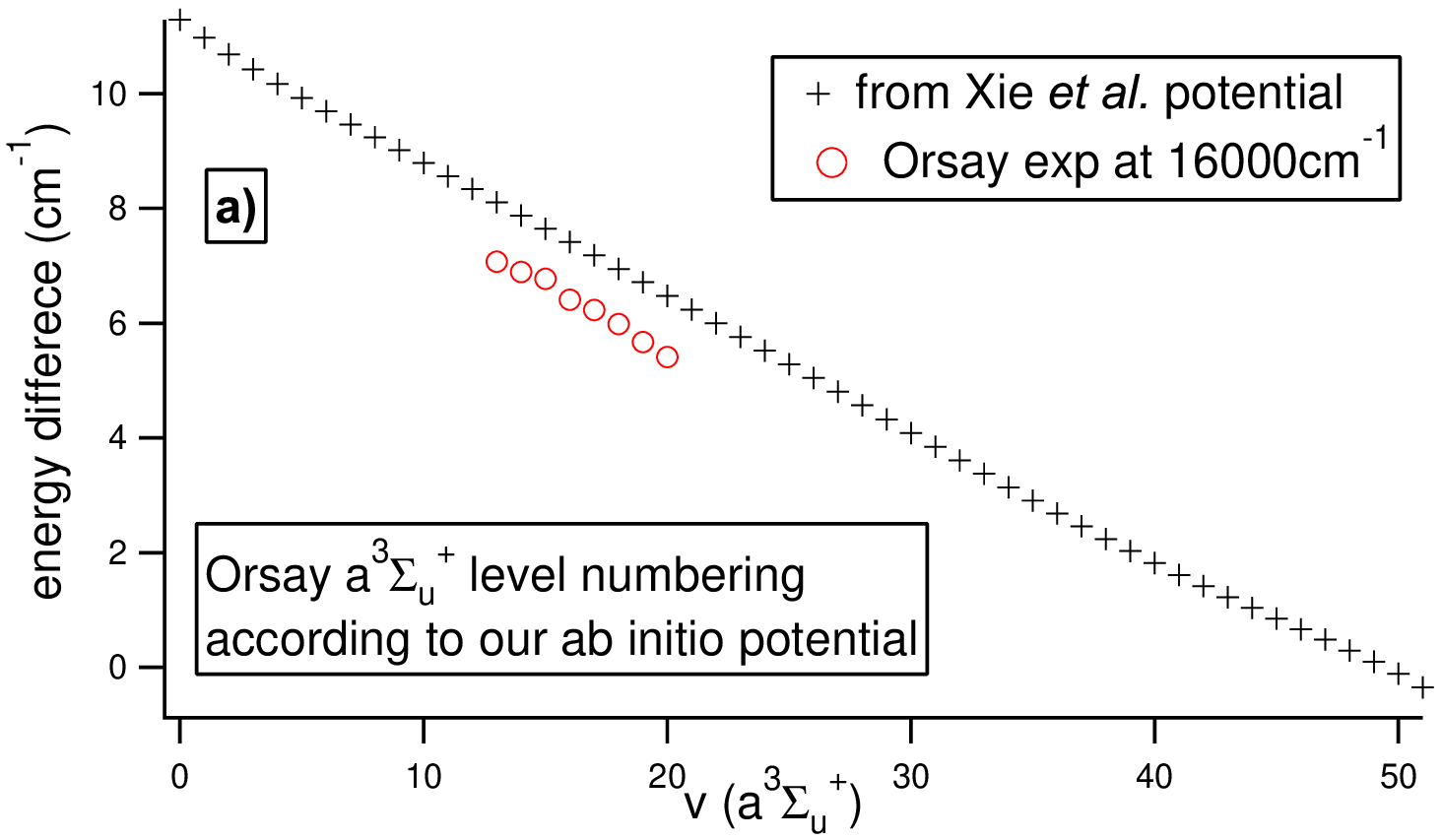}
\includegraphics[width=0.6\textwidth]{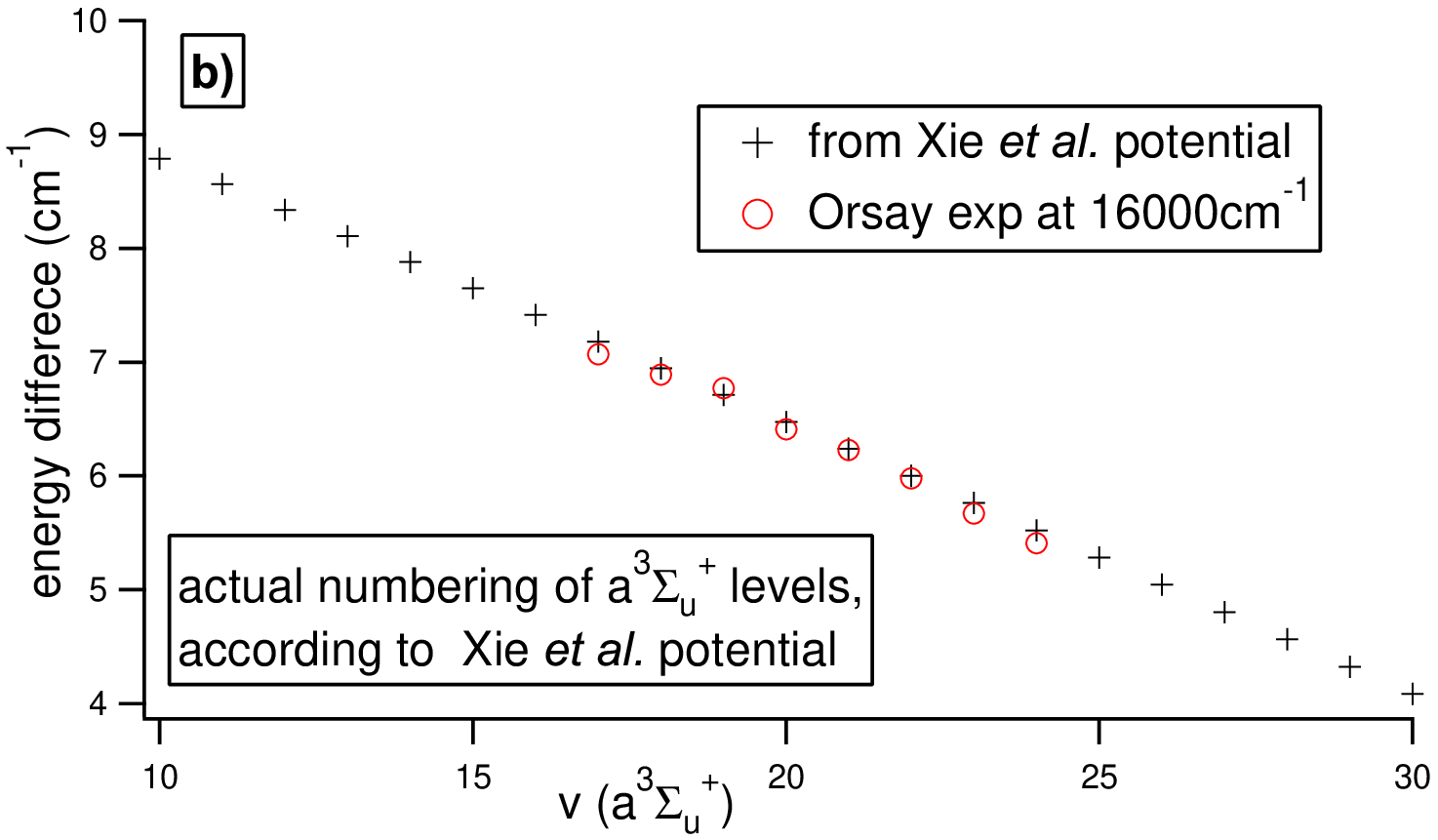}
\caption{(color online) Energy spacings of the $a^3\Sigma_u^+$ vibrational levels $v"$ yielded by Xie {\it et al.}~\cite{xie:051102} (crosses). Open circles represent the energy spacing of the $a^3\Sigma_u^+$ levels deduced from the line positions reported in Table \ref{tab:16000_G1} (see text): (a) levels are labeled according to the numbering of the present theoretical curve; (b) level numbering is shifted upwards by four units.}
\label{fig:spacings_a3su}
\end{figure}

The energy spacings between the successive lines of a given $v'$ sequence give the level spacing of the $a^3\Sigma_u^+$ state, with a level numbering depending on the chosen potential curve used to model the spectra. Fig.~\ref{fig:spacings_a3su}a shows that the energy spacing of the relevant $a^3\Sigma_u^+$ levels decreases linearly with $v"$ with the same slope as the one measured in ref.~\cite{xie:051102}, but with a clear shift in the vibrational assignment. A \emph{local} shift of the vibrational numbering $v"$ by four units yields a good agreement with ref.~\cite{xie:051102} (Fig.~\ref{fig:spacings_a3su}b). This result indicates that the theoretical $a^3\Sigma_u^+$ potential curve has a correct overall shape, but its depth has to be slightly adjusted to match the vibrational numbering of ref.~\cite{xie:051102} in the observed range. Relying on the level numbering vibrational levels $v"$ of the experimental $a^3\Sigma_u^+$ potential curve~\cite{xie:051102}, we assigned many lines which are listed in Table~\ref{tab:16000_G1}. Levels $v"$ in the 17-25 range are populated, which indeed correspond to the prediction of Fig.~\ref{fig:vibpop_astate}b above. In the next sections, we will rely on the experimental $a^3\Sigma_u^+$ potential curve to assign the recorded lines.

\subsection{Ionization through the $(1)^3\Sigma_g^+(6s+6p)$ state}
\label{1_3_Sigma}

\begin{figure}
\centering
\includegraphics[width=0.6\textwidth]{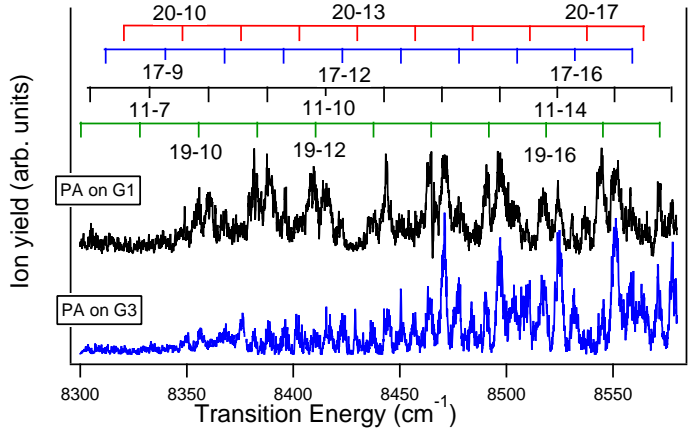}
\caption{The portion between 8300 and 8600~cm$^{-1}$ of the experimental photoionization spectrum recorded in Pisa, with the PA laser frequency tuned on the giant lines $G1$ and $G3$. The transitions are labeled with vibrational indexes $(v",v')$ corresponding to the $a^3\Sigma_u^+ (v") \rightarrow(1)^3\Sigma_g^+ (v')$ transitions, according to the first assignment in Table~\ref{tab:tab_8000}. Four series with $v"=11, 17, 19, 20$ are identified.}
\label{fig:Pisa_8000}
\end{figure}

The REMPI ionization path through the $(1)^3\Sigma_g^+(6s+6p)$ state is neither the easiest scheme to apply nor the first one discovered, but it involves an intermediate state with the lowest possible energy for the first step of the REMPI transition. The corresponding laser has been scanned in the 8000-9500~cm$^{-1}$ range with the Pisa setup.  The two-color detection scheme described in Sec.~\ref{Pisa_setup} is necessary because the absorption of two identical infrared photons is not energetically sufficient to ionize the molecule, which requires $\sim$~25500~cm$^{-1}$ energy above the $6s+6s$ limit. Thus the IR photon energy is scanned along the $a^3\Sigma_u^+$-$(1)^3\Sigma_g^+$ band, while a frequency-fixed green photon photoionizes the excited cesium dimers. Molecular ions are detected for most photon energies between 8350 and 9000~cm$^{-1}$. We display in Fig.~\ref{fig:Pisa_8000} a portion of the recorded spectra between 8300 and 8600~cm$^{-1}$. The relatively low signal-to-noise ratio is due to the rather low and fluctuating laser power obtained with the Raman cell method, and to the large line width of the pulsed dye laser ($\sim 1.8$~cm$^{-1}$). The horizontal scale of the absolute transition energy is obtained by measuring the ionization laser wavelength with a monochromator (Jobin Yvon HR1000), which is calibrated by detecting the ion yield in correspondence of some atomic cesium resonant lines.

The photoionization spectra show quite broad lines, whose width varies in the 2-5~cm$^{-1}$ range. While the minimum width can be explained on the basis of the laser line width, the larger features can be interpreted as the signature of the coalescence of nearby lines. Tuning the PA laser on the $G1$ or $G3$ lines yields similar spectra, as shown in Fig.~\ref{fig:Pisa_8000}. Again, this is an indication that the $G1$ and $G3$ resonances are indeed induced by the same mechanism.

Most of the main observed lines can be assigned with a suitable modeling of the $(1)^3\Sigma_g^+$ electronic state, provided by the high-resolution Fourier transform spectroscopic study of ref.~\cite{Amiot85} performed in a hot cesium vapor. A large number of rovibrational levels spanning the transition energy range between 12000 and 14300~cm$^{-1}$ from the bottom of the $X^1\Sigma_g^+$ ground state potential curve have been assigned, overlapping the range explored in the present work. Amiot and Verg\`es~\cite{Amiot85} modeled the set of measured transitions with a Dunham expansion~\cite{dunham} that two of us used recently to assign the lines observed in the ultracold molecule experiment of ref.~\cite{danzl2009}. This latter work unambiguously determined the energy spacing between the uppermost $a^3\Sigma_u^+$ vibrational level (\emph{i.e.} almost the position of the $6s+6s$ dissociation limit) and the lowest $(1)^3\Sigma_g^+$ levels $v'=31-35$ observed in ref.~\cite{Amiot85}.

A tentative assignment of the observed lines is reported in Table~\ref{tab:tab_8000}. Due to their somewhat large width, most of the lines can be assigned to two possible transitions with similar accuracy. Following the interpretation of Section~\ref{3_3_Sigma}, these lines involve $a^3\Sigma_u^+$ levels $v"=17, 19, 20$ belonging to the range of populated levels predicted by the model of Fig.~\ref{fig:vibpop_astate}b. However several transitions from $v"=11$ seem also to be observed.

We remark also in Fig.~\ref{fig:Pisa_8000} that even if the line intensities cannot be fully trusted due to strong saturation, the signal vanishes for transition energies lower than 8330~cm$^{-1}$, which can be compared to the energy difference between the minimum of the $(1)^3\Sigma_g^+$ state from ref.~\cite{Amiot85} and the $a^3\Sigma_u^+$ experimental curve of ref.~\cite{xie:051102}, $T_e=8232(1)$~cm$^{-1}$. Actual calculation of FC coefficients between the $(1)^3\Sigma_g^+$ and the $a^3\Sigma_u^+$ states only partially supports this experimental observation.

\subsection{Ionization through the $(2)^3\Sigma_g^+(6s+5d)$ and the $(2)^3\Pi_g(6s+5d)$ states}
\label{Sigma_Pi}

The range of laser energies for the first REMPI step between 13200 and 14300~cm$^{-1}$ was actually the one used in the earlier experiments in Orsay~\cite{dion2002}. It was proved to be efficient for the detection of ultracold molecules created in the $(a)^3\Sigma_u^+$ state, through one-photon resonance with levels of the $(2)^3\Sigma_g^+(6s+5d)$ or the $(2)^3\Pi_g(6s+5d)$ states.

\begin{figure}
\centering
\includegraphics[width=0.6\textwidth]{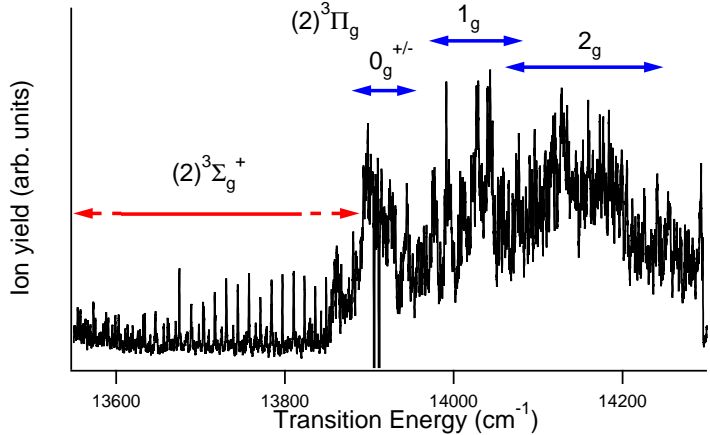}
\caption{Ionization spectrum recorded in Orsay with PA laser tuned on the $G3$ resonance and the pulsed lasers with LDS722 dye. As in ref.~\cite{dion2002}, we can identify the region where ionization proceeds via the $(2)^3\Sigma_g^+(6s+5d)$ only (below 13850~cm$^{-1}$), or via both this latter state and the three fine structure components 0$_{\rm g}^{-/+}$, 1$_{g}$ and 2$_{g}$ of the $(2)^3\Pi_g(6s+5d)$ state.}
\label{fig:LAC_13000_G3}
\end{figure}

In Figure~\ref{fig:LAC_13000_G3}, we show the ionization spectrum recorded in Orsay when the PA laser is tuned on the $G3$ resonance, and the ionization achieved by the pulsed laser with LDS722 dye. A similar spectrum is recorded when the PA laser is tuned on $G1$. These spectra extend from 13500~cm$^{-1}$ to 14300~cm$^{-1}$. Below 13850~cm$^{-1}$ the well-resolved lines correspond to transitions assigned to the $a^3\Sigma_u^+$ -$(2)^3\Sigma_g^+(6s+5d)$ transitions. Above 13850~cm$^{-1}$ the lines are not resolved anymore as the spectrum becomes more complex. We can clearly distinguish the three bands corresponding to the three fine-structure components 0$_{\rm g}^{-/+}$, 1$_{g}$ and 2$_{g}$ of the $(2)^3\Pi_g (6s+5d)$ state. The system $a^3\Sigma_u^+$-$(2)^3\Pi_g (6s+5d)$ was experimentally investigated using laser-induced fluorescence and resonant two-photon ionization~\cite{diemer1991}. The $T_e$ values which give the energy position of the bottom of the potential curves relative to the bottom of the $a^3\Sigma_u^+$ state were determined for each component. For the lowest component  $\Omega=0^+$ this energy was found to be at 13928(1)~cm$^{-1}$ in ref.~\cite{diemer1991}, \emph{i.e.} at 13648(2)~cm$^{-1}$ above the $6s+6s$ dissociation limit. This is consistent with the spectrum of Figure~\ref{fig:LAC_13000_G3} which shows that the lowest band of the $a^3\Sigma_u^+$ - $(2)^3\Pi_g (6s+5d)$ rises up between these two values.

\begin{figure}
\centering
\includegraphics[width=0.8\textwidth]{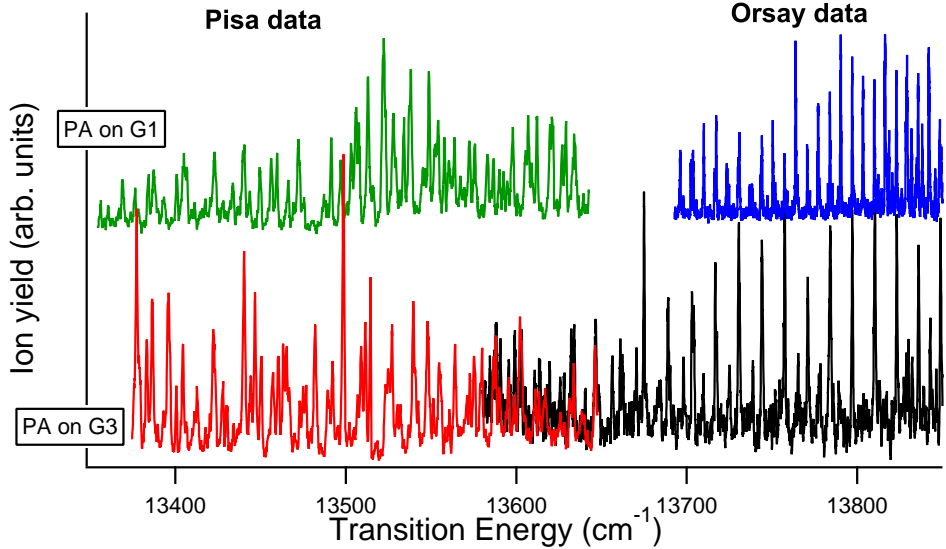}
\caption{(color online) Region of the ionization spectrum involving the intermediate $(2)^3\Sigma_g^+(6s+5d)$ state, recorded in Orsay and in Pisa. The PA laser was tuned to the $G1$ (upper traces) and $G3$ (lower traces) resonances respectively.}
\label{fig:13000_complet}
\end{figure}

We will concentrate on the resolved parts of the spectra recorded in Orsay and in Pisa in the region 13350~cm$^{-1}$to 13850~cm$^{-1}$ for the first REMPI excitation laser (the LDS751 dye in Pisa, and the LDS722 dye in Orsay), when the PA laser is tuned either on the $G1$ or on $G3$ resonances. The entire data set is displayed in Fig.~\ref{fig:13000_complet}, showing that spectra obtained in Pisa and Orsay nicely overlap, and that line patterns are again similar for both $G1$ and $G3$ excitations.

\begin{figure}[htbp]
\centering
\includegraphics*[width=0.8\textwidth]{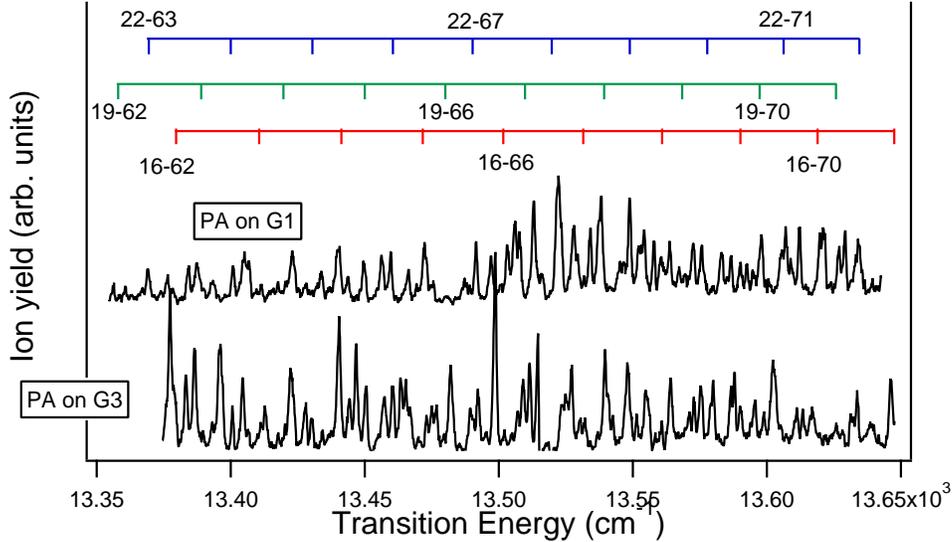}
 \caption{Part of the experimental photoionization spectrum registered in Pisa between 13350 and 13650~cm$^{-1}$. The PA laser was tuned to the $G1$ (upper traces) and $G3$ (lower traces) resonances respectively. Energies for assigned transitions are reported in Table~\ref{tab:13000_Pisa}.}
\label{fig:2_3Sigma_g_Pisa}
\end{figure}

\begin{figure}[htbp]
\centering
\includegraphics[width=0.8\textwidth]{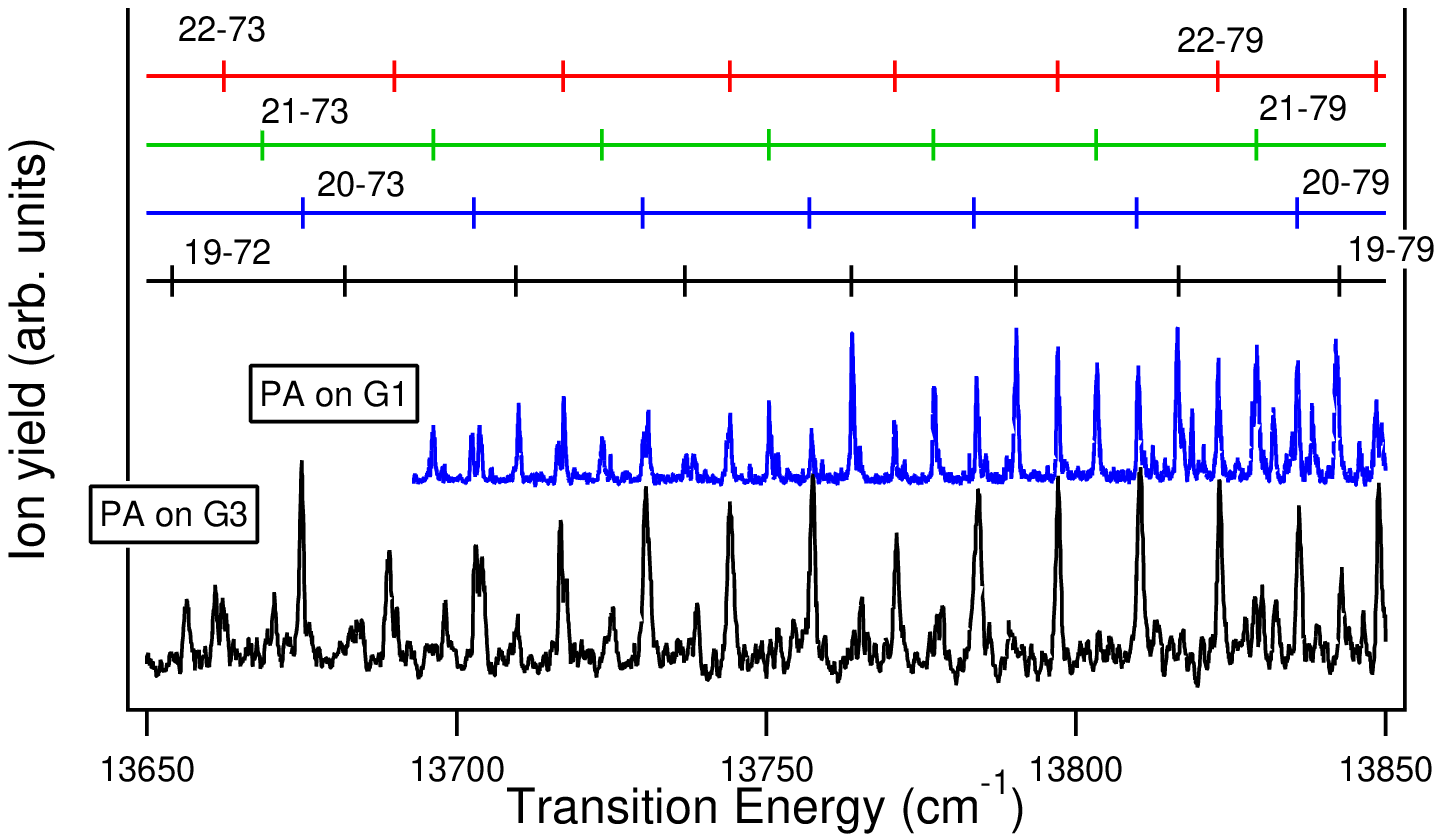}
\caption{Part of the experimental photoionization spectrum registered in Orsay between 13650 and 13850~cm$^{-1}$. The PA laser was tuned to the $G1$ (upper traces) and $G3$ (lower traces) resonances respectively. Energies for assigned transitions are reported in Table~\ref{tab:13000_LAC}.}
\label{fig:2_3Sigma_g_LAC}
\end{figure}

The lines are assigned to vibrational transitions of the $a^3\Sigma_u^+$-$(2)^3\Sigma_g^+(6s+5d)$ system by comparing the experimental spectra to theoretical ones as above (Figures~\ref{fig:2_3Sigma_g_Pisa} and~\ref{fig:2_3Sigma_g_LAC}). For the excited state $(2)^3\Sigma_g^+(6s+5d)$ we used an \emph{ab initio} potential curve from ref.~\cite{aymar2010}, which had to be shifted upward by about 220.0~cm$^{-1}$ to obtain satisfactory agreement with experimental lines. We thus take for this state the energy $T_e$=11172(1)~cm$^{-1}$  relative to the bottom of the experimentally determined $a^3\Sigma_u^+$ potential curve. This demonstrates that the shape of the chosen $(2)^3\Sigma_g^+(6s+5d)$ is correct in the energy region relevant for this study. The analysis indicates that the $a^3\Sigma_u^+$ levels populated by spontaneous decay of the $G1$ and $G3$ lines fall within the predicted range of Fig.~\ref{fig:vibpop_astate}b. The measured energy for each line and its vibrational assignment are given in Tables~\ref{tab:13000_Pisa} and~\ref{tab:13000_LAC} for the Pisa and Orsay spectra respectively, as well as calculated transition energies including the upward shift of 220.0~cm$^{-1}$.

\section{Conclusions}
\label{conclusions}

In this collaborative work between the Orsay and Pisa teams, we have presented experimental spectra for the REMPI detection of ultracold Cs$_{2}$ molecules produced
in a magneto-optical trap by photoassociation, with their analysis relying on previous spectroscopic data, or up to date theoretical determination of the involved Cs$_{2}$ molecular states. The PA step is performed on two of the so-called \emph{giant lines}, \emph{i.e.} three intense PA resonances detected in the PA spectrum of the excited $0_g^-(6s+6p_{3/2})$ double-well state, whose levels are known to spontaneously decay into vibrational levels of the stable lowest triplet state $a^3\Sigma_u^+$ of Cs$_{2}$. Using three different pathways for the REMPI excitation, \emph{i.e.} involving the three excited molecular states $(1)^3\Sigma_g^+ (6s+6p)$, $(2)^3\Sigma_g^+ (6s+5d)$, and $(3)^3\Sigma_g^+(6s+7s)$, we were able to interpret the spectra in a way which confirms the model of Vatasescu \emph{et al.}~\cite{vatasescu2006} for the mechanism responsible for two of these giant lines, labeled as $G1$ and $G3$: they arise from the tunneling of the $0_g^-(6s+6p_{3/2})$ vibrational wave functions through the potential barrier separating the two wells of the corresponding potential curves. This pattern induces spontaneous decay towards deeply-bound levels of the $a^3\Sigma_u^+$ state, mainly in the $v"=15-24$ range. Such a state-selective ionization of ground-state molecules is a promising tool for optical pumping of cold molecules into more deeply-bound vibrational levels, and generally for investigations of level-dependent molecular properties. Therefore, the formation of ultracold Cs$_2$ molecules via PA of such tunneling resonances is a promising way for creating stable $v"=0$ molecules using a further simple population transfer for instance with a STIRAP scheme.

The present study also yielded new spectroscopic information on Cs$_2$ molecular states with triplet multiplicity (Table \ref{tab:spectrodata}), which are not easily accessible by conventional high-resolution spectroscopy in thermal vapors. In particular, we were able to determine the absolute energy position of the previously unobserved $(2)^3\Sigma_g^+ (6s+5d)$ and $(3)^3\Sigma_g^+(6s+7s)$ potential curves. The obtained data enrich the knowledge of the still poorly-known triplet manifolds, and also represent a strong constraint for checking the accuracy of \emph{ab initio} calculations.

\begin{acknowledgments}
This work is supported in Orsay by the "Institut Francilien de Recherche sur les Atomes Froids" (IFRAF). This work is also partially supported by the network "Quantum Dipolar Molecules (QuDipMol)" of the EUROQUAM framework of the European Science Foundation (ESF). A.F.  has been supported by the "\emph{Triangle de la Physique}" under
contracts 2007-n.74T and 2009-035T "GULFSTREAM".  M.A. thanks the EC-Network EMALI, the "\emph{Universit\'e Franco-Italienne}" (Galileo Project) and the partnership between University of Paris-Sud and the University of Pisa (COGITO).
\end{acknowledgments}

\newpage
\bibliography{Pisa_Orsay_10}

\begin{thebibliography}{10}
\bibitem{dunham}%
  \BibitemOpen
  \bibinfo {note} {The $Y_{0,0}=T_e$ Dunham coefficient gives the energy
  position of the bottom of an excited-state potential curve with respect to
  the bottom of the $X^1\Sigma_g^+$ ground state. As the depth $D_e$ of this
  state is spectroscopically known with high accuracy, it is thus possible to
  deduce the energy postion of the excited state with respect to the $s+s$
  dissociationi limit.}%
  \bibAnnoteFile{Stop}{dunham}%
\end{thebibliography}

\newpage

\section*{Tables}

\begin{table}[htbp]
\caption{Recorded transition energies (with the Orsay setup) between $a^3\Sigma_u^+$ vibrational levels $v"$ and $(3)^3\Sigma_g^+(6s+7s)$ vibrational levels $v'$, when the PA laser is tuned on the $G1$ resonance. The vibrational numbering for the $a^3\Sigma_u^+$ levels is the one provided by the experimental potential.}
\label{tab:16000_G1}
\end{table}

\begin{table}[htbp]
\caption{Recorded transition energies (with the Pisa setup) between $a^3\Sigma_u^+$ vibrational levels $(v")$ and $(1)^3\Sigma_g^+ (6s+6p)$ vibrational levels $v'$ when the PA laser is tuned on the $G3$ resonance. Two possible assignments have been determined with similar accuracy. The vibrational numbering for the $a^3\Sigma_u^+$ levels is the one provided the experimental potential.}
\label{tab:tab_8000}
\end{table}

\begin{table}[htbp]
\caption{Recorded transition energies (with the Pisa setup) between $a^3\Sigma_u^+$ vibrational levels $(v")$ and $(2)^3\Sigma_g^+ (6s+5d)$ vibrational levels $v'$ when the PA laser is tuned on the $G1$ and $G3$ resonances. Calculated transition energies including a shift of 220.05~cm$^{-1}$ are also reported (see text). Overlapping lines, indicated by indexes (1) to (7), Overlapping lines, indicated by indexes (1) to (7), are lines whose assignment is ambiguous and could be twofold. The vibrational numbering for the $a^3\Sigma_u^+$ levels is the one provided by the experimental potential}
\label{tab:13000_Pisa}
\end{table}

\begin{table}[htbp]
\caption{Recorded transition energies (with the Orsay setup) between $a^3\Sigma_u^+$ vibrational levels $(v")$ and $(2)^3\Sigma_g^+ (6s+5d)$ vibrational levels $v'$ when the PA laser is tuned on the $G1$ and $G3$ resonances. Calculated transition energies including a shift of 220.05~cm$^{-1}$ are also reported (see text). The vibrational numbering for the $a^3\Sigma_u^+$ levels is the one provided by the experimental potential.}
\label{tab:13000_LAC}
\end{table}

\begin{table}[htbp]
\caption{Summary of the spectroscopic constants of the various excited triplet states of Cs$_2$ deduced from the present study. For clarity, the spectroscopic constants of the theoretical and experimental $a^3\Sigma_u^+$ state are also given in this table. The $T_e$ values are determined with respect to the bottom of the experimental $a^3\Sigma_u^+$ potential curve of ref.~\cite{xie:051102}. Therefore, the $T_e$ of the theoretical $a^3\Sigma_u^+$ curve indicates that its depth is smaller by about 40~cm$^{-1}$ than the one of ref.~\cite{xie:051102}. The experimental value for the harmonic constant of the $(2)^3\Sigma_g^+ (6s+5d)$ state is not reported because the accessible range of vibrational levels lies in the anharmonic part of the potential curve.}
\label{tab:spectrodata}
\end{table}

\end{document}